\documentclass[journal]{IEEEtran}
\usepackage{cite}

\ifCLASSINFOpdf
   \usepackage[pdftex]{graphicx}
\else
\fi
\usepackage{amsmath,amssymb,amsfonts}
\usepackage{algorithm, algpseudocode}
\usepackage{multirow}

\begin{document}
%
\title{MILP-based optimal day-ahead scheduling for system-centric CEMS supporting different types of homes and energy trading}
%
%
%

\author{Huy~Truong~Dinh,~\IEEEmembership{Student Member,~IEEE,}
		Dongwan Kim,~\IEEEmembership{Member,~IEEE}
        and~Daehee~Kim,~\IEEEmembership{Member,~IEEE}
\thanks{This work was supported by the Korea Institute of Energy Technology Evaluation and Planning (KETEP) and the Ministry of Trade, Industry \& Energy (MOTIE) of the Republic of Korea (No. 20184030202130). This work was also supported by the Soonchunhyang University Research Fund.}
\thanks{Huy Truong Dinh and Daehee Kim are with the Department of Future Convergence Technology, Soonchunhyang University, Asan 31538, South Korea (e-mail: tdhuy@sch.ac.kr, daeheekim@sch.ac.kr). Dongwan Kim is with the Department of Electronics Engineering, Dong-A University, Busan 49315, South Korea}
}

%
%

\markboth{Journal of \LaTeX\ Class Files,~Vol.~14, No.~8, August~2015}%
{Shell \MakeLowercase{\textit{et al.}}: Bare Demo of IEEEtran.cls for IEEE Journals}
%



\maketitle

\begin{abstract}
Optimal day-ahead scheduling for a system-centric community energy management system (CEMS) is proposed to provide economic benefits and user comfort of energy management at the community level. Our proposed community includes different types of homes and allows prosumers to trade energy locally using mid-market rate pricing. A mathematical model of the community is constructed and the optimization problem of this model is transformed into an MILP problem that can be solved in a short time. By solving this MILP problem, the optimization of the overall energy cost of the community and satisfaction of the thermal comfort at every home are achieved. For comparison, we also establish two different scenarios for the same community: a prosumer-centric CEMS and no CEMS. The simulation results demonstrate that the overall energy cost of the community with the system-centric CEMS is the smallest among the three scenarios and is only half that of the community with the prosumer-centric CEMS. Moreover, by using linear transformation, the computational time of the optimization problem of the proposed system-centric CEMS is only 118.2 s for a 500-home community, which is a short time for day-ahead scheduling of a community. 
\end{abstract}


%
\IEEEpeerreviewmaketitle

\section{Introduction}

\IEEEPARstart{W}{ith} the rapid increase in many distributed energy resources (DERs), such as renewable energy system (RESs) and energy storage systems (ESSs) at many smart homes, community energy management system (CEMSs) for local energy communities (LECs) to reduce the energy cost of this group and increase the benefit for each member have received substantial research attention. A CEMS generally includes two parts: energy management for its DERs and energy trading for its members. The approaches of a CEMS can mainly be divided into two categories depending on the management of the DERs: the prosumer-centric and system-centric approaches \cite{ye2021scalable}. The first category allows prosumers to maintain complete control of their DERs. However, the integration of the information of the trading model into the scheduling model of the prosumer devices at each home is a significant challenge. Hence, this approach is usually divided into two stages. In the first stage, each prosumer schedules their DERs to optimize the individual energy cost without information of the trading model. In the second stage, prosumers join the local market to improve their benefits \cite{cui2020new}, \cite{lilla2019day}, \cite{lin2018two} \cite{cui2019peer}, \cite{orozco2020multistage}, \cite{cui2020game}. The second category focuses on the construction of a central entity that collects the information of all DERs inside the community, and schedules them to optimize its overall objectives and maximally match the electricity demand of consumers with the electricity generation of producers within the community \cite{wang2019incentivizing}, \cite{li2021energy}, \cite{nan2018optimal}. 

The local energy market (LEM) of an LEC is an important element that makes a substantial contribution to the success of the LEC. In particular, the LEM enables energy sharing among prosumers, thereby offering a powerful and complete exploitation of the prosumer DERs within the community and reducing the grid dependence of the community. It also helps to increase the benefits for prosumers because the pricing proposed by the LEM is usually better than that of the electricity provider. As the conventional retail market is not suitable for the LEM, a new market has been studied in many previous studies to coordinate prosumers efficiently. Depending on the degree of decentralization, LEM structures can be classified into fully peer-to-peer (P2P) market and community-based (CB) market \cite{crespo2020community}. In P2P-LEM, there is no need for a central entity; prosumers are directly interconnected and trade bilaterally with one another. The main advantage of P2P-LEM is that it allows prosumers to have independent control of their trading \cite{van2020integrated}. The well-known pricing schemes that are generally used in this type of market are the alternating direction method of multipliers (ADMM) \cite{morstyn2018multiclass}, \cite{nguyen2020optimal} and game theory \cite{paudel2018peer}, \cite{tushar2020coalition}. In the CB-LEM, there is a central entity that collects information of the trading and schedules of each prosumer and optimizes the overall profits in the market \cite{khorasany2020framework}, \cite{zhou2020smart}. The interest of the community is predominant in this market, and certain prosumers may sacrifice their profits for the overall profit of the community. The mid-market rate (MMR) \cite{long2017peer}, \cite{li2020computationally} and auction-based \cite{yu2021continuous}, \cite{doan2021peer} pricing schemes are popular in this type of market.

Based on the above background, we focus on investigating an optimization problem for a system-centric CEMS to minimize the overall energy cost of an LEC that supports different types of homes and the CB-LEM between these. The major challenge of the system-centric approach is the computational time that is required to solve an optimization problem containing a massive number of variables and constraints. However, our problem is transformed into a mixed integer linear programming (MILP) problem using linear transformation, which can be solved in an acceptable time. Specifically, this study makes the following contributions:
\begin{itemize}
\item We propose a community that includes four types of homes: homes with an RES and ESS, homes with an RES only, homes with an ESS only, and normal homes without an RES or ESS. The community supports a CB-LEM to use the RESs and ESSs of the homes within the community fully. The RES energy of a home is used in many manners: for home devices, for storage in the ESS of the home, or to sell to the community. Likewise, the ESS energy of a home is used for the home load or for selling to the community. The ESS is also used to store RES energy or cheap energy from the community.
\item With the above functionalities of the RES and ESS, we construct general mathematical formulas of the DERs and load for each home and for the entire community. Based on these formulas, an optimization problem of the community is built and successfully transformed into an MILP problem that can be solved in a short time. The optimization of the daily energy cost of the community and satisfaction of thermal comfort are achieved by solving this problem. The simulation results demonstrate that the daily energy cost of the community with a system-centric CEMS is only approximately half that of the community with a prosumer-centric CEMS.
\item Optimization problems of many communities with different numbers of homes are built and solved to evaluate the computational time of the proposed community. The simulation results indicate that the computational time of the community with 500 homes is only 118.2 s, which is a short time for day-ahead scheduling.
\item We also demonstrate the trade-off of the MMR pricing in the local energy trading of the community. This trade-off allows the benefits of different types of homes to be flexibly changed.
\end{itemize}

The remainder of this paper is organized as follows: Section II provides a review of the related works. In Section III, the construction and transformation of our system model and the optimization problem are presented in detail. Section IV describes the local energy trading mechanism with MMR pricing. In Section V, the simulations and results are discussed. Finally, Section VI outlines the conclusions and potential future works.

%
%
%
%

\section{Related Works}
Numerous studies have been conducted on CEMSs. In this section, because our proposed method is day-ahead scheduling, we focus on reviewing existing day-ahead scheduling methods for an LEC.

In \cite{nan2018optimal}, the authors proposed a residential community in which a load aggregator was used to collect all information of the DERs and loads at each home, and to construct an optimization problem to minimize the energy cost of the community. By solving this problem using the CPLEX solver, this load aggregator provided a day-ahead schedule for all DERs and appliances within the community. Every home of the community had its own RES and ESS. However, in this study, the authors assumed that solar power was only used for the community load, and it was not sold to the grid or stored in the ESS. Moreover, local energy trading was not supported in this community.

The authors of \cite{hossain2019energy} proposed day-ahead energy management for a community of 15 houses to minimize its energy cost. There were only three DERs that were shared among all members within the community: wind generation, solar photovoltaic (PV) generation, and an ESS. The RES energy could be used for the local load, stored in the ESS, or sold to the grid. The optimization problem of the community was solved by the particle swarm optimization (PSO) algorithm and a day-ahead schedule of sharing the DERs was provided. However, the individual DERs at each home were not considered.

In \cite{zhou2020smart}, a community with an EMS, including a local P2P market and a user-dominated demand side response (UDDSR) program, was proposed to reduce the energy bills of the community. Certain homes within the community only had a PV system as the RES, and a central sharing ESS was integrated into the community to store surplus energy from its members. The UDDSR program collected flexible bids from the homes, including information of their controllable appliances. The program solved the MILP optimization problem and the schedules of controllable appliances were provided to maximize the load balancing between homes using the GUROBI solver. However, individual ESSs were not considered in this study.

In \cite{lilla2019day}, day-ahead scheduling of a prosumer-centric LEC using ADMM was proposed. In this study, each home had an individual PV system as the RES and an individual ESS. An optimization function to minimize the energy cost and power loss for each prosumer was built. Subsequently, the prosumers cooperated to solve the problem distributively by using ADMM algorithms. However, the RES energy was only used for home load and selling at each home; it could not be stored in the ESS. Moreover, this study only supported homes that included both an RES and ESS, and controllable load scheduling to satisfy user comfort was not considered.

In \cite{cui2020new}, the authors proposed an energy sharing framework for a community in which each building had its own DERs, including an RES and ESS. Each building first optimized its DERs and controllable load to minimize the energy cost without energy sharing. Thereafter, a non-cooperative sharing game was used to determine the energy sharing profile and corresponding payments of each building. However, in this study, the RES energy could not be stored in the ESS, and homes that had an RES only or ESS only were not considered.

The authors of \cite{nizami2019multiagent} proposed a two-stage EMS for a group of buildings. In the day-ahead stage, an MILP-based objective function of the scheduling model was constructed to minimize the energy cost while maintaining the user comfort in each building. Subsequently, in the real-time stage, each building participated in a transactive market to maximize the profits. However, in the day-ahead scheduling model, no energy was sold to the public grid and any excess energy was stored in the ESS.

Motivated by the above works, we propose a community with a system-centric CEMS that supports different types of homes with individual DERs: homes with both an RES and ESS, homes with an individual RES only or individual ESS only, and homes without an RES or ESS. Moreover, individual RES energy can be used in many manners: for home loads, for selling to outside the community, or for charging to the individual ESS of a home if this home has its own ESS. In our community, local energy trading between prosumers is also supported using the MMR pricing scheme.

\section{System Model and Problem Formulation}
\begin{figure}[]
\centering
\includegraphics[scale=0.4]{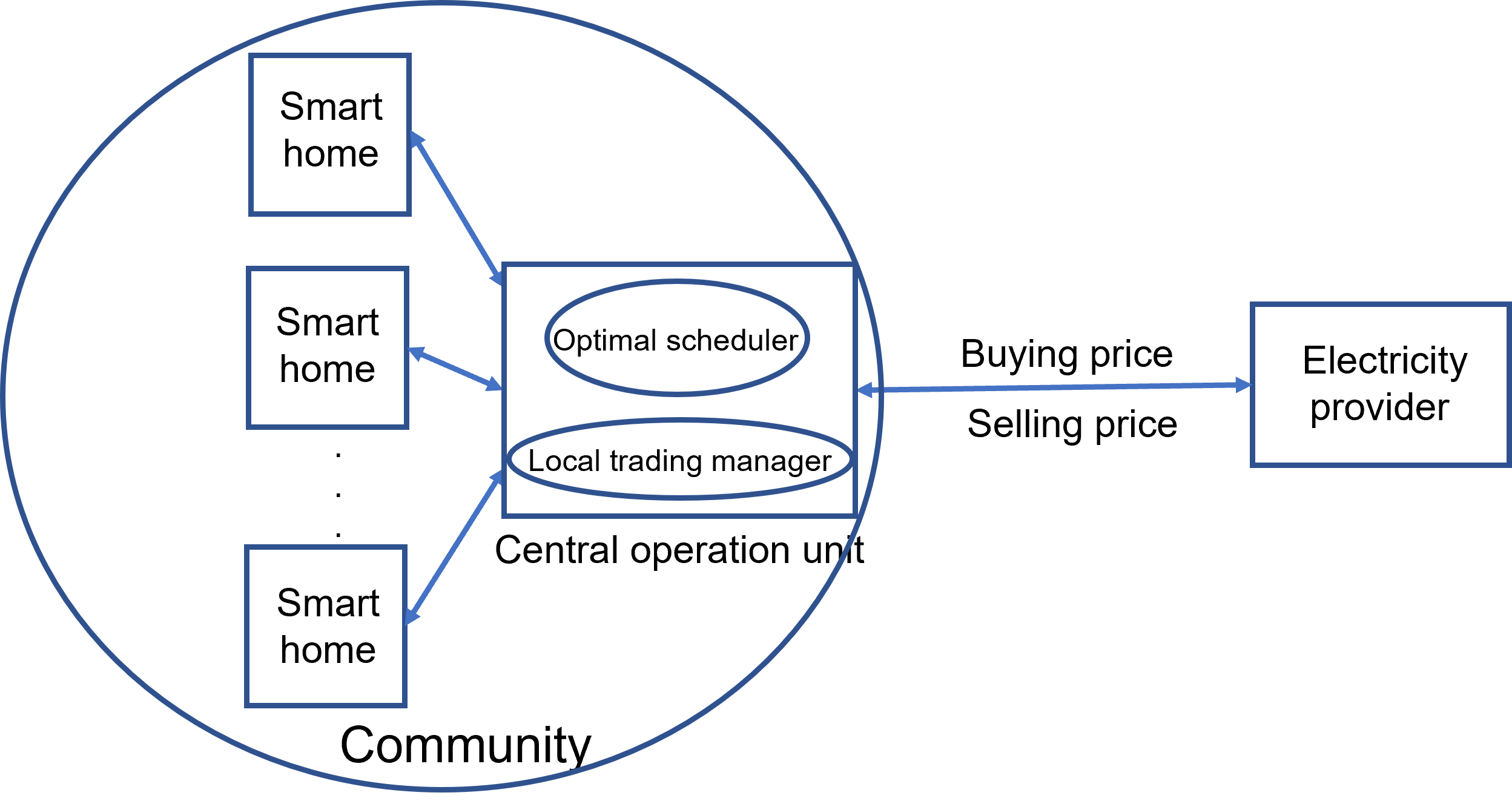}
\caption{Structure of community energy management.}
\label{community}
\end{figure}

In this section, we consider a community including a group of smart homes $H=\{1,2,...,N\}$, as illustrated in Fig. \ref{community}. Within the community, the central operation unit collects information from all homes and from the electricity provider (EP), such as the DERs of each home, price information, forecast temperature, and solar irradiation. An optimal scheduler (OS) that includes an optimization algorithm and local trading manager that is responsible for coordinating the local trading within the community are installed inside the central operation unit. At the beginning of the day, useful information is received from the homes and EP. Subsequently, the OS is run to create an optimized schedule for all DERs and devices of each home during a day, whereas the local energy trading manager is run to calculate the daily energy cost of each home based on its local buying/selling prices.

\begin{figure}[]
\centering
\includegraphics[scale=0.4]{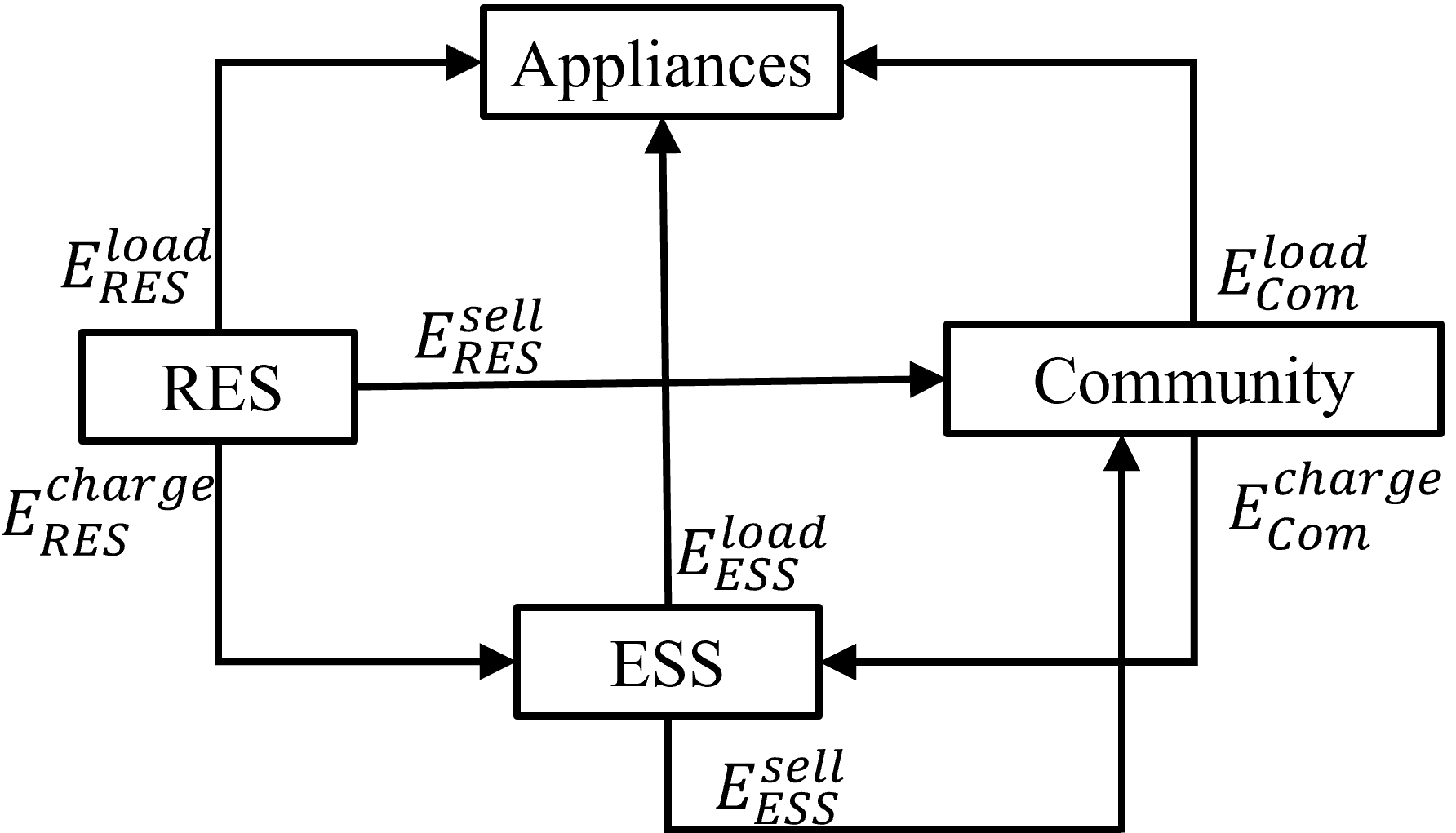}
\caption{Energy flows in smart home with RES and ESS.}
\label{electricity_flow}
\end{figure}

Each home may have both an ESS and RES, only one of these, or neither. The energy flows of a home, which include both the ESS and RES, in the community are depicted in Fig. \ref{electricity_flow}. If a home has an RES, its energy can be used for home loads, ESS charging, or selling to the community. If a home has an ESS, its energy is used for home loads and selling to the community. The ESS is also used to store electricity from the community at a low price and to provide electricity for home loads at the high price time. We consider two different loads in a home for each time slot: the fixed load and controllable load. Fixed loads are the loads of devices for which the power consumption cannot be changed in this time slot, whereas controllable loads are the loads of devices for which the power consumption can be changed to satisfy certain constraints based on the environmental status. 

We build mathematical formulas and constraints for all RESs, ESSs, and loads in the community during a day from $0$ AM to 12 PM to optimize the daily energy cost of the community. We also divide a day into $T=24$ time slots, and the duration of each time slot is $\Delta t = 1$ h.

\subsection{Controllable Load}
In this study, an adjusted HVAC system in heating mode is considered as the controllable load in a home. Let $p_{i}(t)$ be the input power of the HVAC system in time slot $t$ at home $i$. This variable can be adjusted continuously within a certain range \cite{yu2017online}.
\begin{equation}
0 \leq p_{i}(t) \leq P_{max},
\label{constraint_power_HVAC}
\end{equation}
where $P_{max}$ is the rating power of the HVAC system, and the HVAC energy $E_{i,HVAC}(t) \geq 0$ that is required in time slot $t$ at home $i$ can be calculated as follows:
\begin{equation}
E_{i,HVAC}(t) = p_{i}(t) \cdot \Delta t.
\end{equation}

According to \cite{yu2017online}, \cite{dinh2021milp}, and \cite{yu2019deep}, the indoor temperature $T_{i}^{in}(t+1)$ of home $i$, which is influenced by the input power of the HVAC system $p_{i}(t)$, the indoor temperature $T_{i}^{in}(t)$, and the outdoor temperature $T^{out}(t)$, is calculated as follows:
\begin{equation}
T_{i}^{in}(t+1) = \varepsilon \cdot T_{i}^{in}(t) + (1- \varepsilon) \cdot (T^{out}(t) + \frac{\eta \cdot p_{i}(t)}{A}),
\end{equation}
where $\varepsilon$ is a constant that describes the system inertia. Furthermore, $\eta$ refers to the thermal conversion efficiency and $A$ is the overall thermal conductivity.

When using an HVAC system, the indoor temperature in each time slot of a day must be guaranteed in a comfortable temperature range $[T_{min}, T_{max}]$ at each home in the community. Hence, we obtain the following constraint:
\begin{equation}
T_{min} \leq T_{i}^{in}(t) \leq T_{max}.
\label{constraint_input_temperature}
\end{equation}

\subsection{ESS Model}
Let $E_{i,ESS}^{Level}(t)$ be the energy level of the ESS of home $i$ after time slot $t$. As described in Fig. \ref{electricity_flow}, with $\forall t, \hspace{0.1cm}1 \leq t \leq T$, we obtain the following formula:
\begin{align}
E_{i,ESS}^{Level}(t) = &E_{i,ESS}^{Level}(t-1) - \Big(E_{i,ESS}^{load}(t) + E_{i,ESS}^{sell}(t) \Big) / \eta^{ESS} \nonumber \\
&+ \Big(E_{i,RES}^{charge}(t) + E_{i,Com}^{charge}(t)\Big) \cdot \eta^{ESS}
\label{ESS_Level}
\end{align}
where $E_{i,ESS}^{load}(t) \geq 0$ is the energy from the ESS of home $i$ that is used for appliances in time slot $t$. $E_{i,ESS}^{sell}(t) \geq 0$ is the energy from the ESS of home $i$ that is used for selling back to the community in time slot $t$. $E_{i,RES}^{charge}(t) \geq 0$ is the energy that is stored in the ESS from the RES of home $i$ in time slot $t$. $E_{i,Com}^{charge}(t) \geq 0$ is the energy that is stored in the ESS of home $i$ from the community in time slot $t$. $\eta^{ESS}$ is the ESS efficiency. 

When using the ESS, the following constraints must be satisfied:
\begin{equation}
EL_{i,min} \leq E_{i,ESS}^{Level}(t) \leq EL_{i,max}
\label{ESS_Level_Constraint}
\end{equation}
\begin{equation}
E_{i,RES}^{charge}(t) + E_{i,Com}^{charge}(t) \leq Ch_{i,rate} \cdot \Delta t \cdot mode_{i,ESS}(t)
\label{ESS_Charge_Constraint}
\end{equation}
\begin{equation}
E_{i,ESS}^{load}(t) + E_{i,ESS}^{sell}(t) \leq Dh_{i,rate} \cdot \Delta t \cdot \big(1 - mode_{i,ESS}(t)\big)
\label{ESS_Discharge_Constraint}
\end{equation}
\begin{equation}
mode_{i,ESS}(t) = 
   \begin{cases}
   1 & \quad \text{if ESS is charged in time slot } t \\
   0 & \quad \text{if ESS is discharged in time slot } t
   \end{cases}
\label{charge_discharge_ESS_mode}
\end{equation}
where $EL_{i,min}$ and $EL_{i,max}$ are the minimum and maximum energy levels of the ESS of home $i$, respectively. $Ch_{i,rate}$ and $Dh_{i,rate}$ are the maximum charge and discharge rates of the ESS, respectively. $mode_{i,ESS}(t)$ is a binary variable to avoid simultaneous ESS charging and discharging in time slot $t$ at home $i$. It is assumed that the ESS cannot be charged and discharged simultaneously.

As we only consider our system during a day (with no net accumulation for the following day), the energy level should be returned to the initial energy level $EL_{i,0}$ at the end of the day. Thus, we have
\begin{equation}
E_{i,ESS}^{Level}(T) = EL_{i,0}.
\label{ESS_last_level_Constraints}
\end{equation}


\subsection{RES Model}
According to \cite{ru2012storage} and \cite{dinh2021optimal}, the output energy $E_{i,RES}(t) \geq 0$ from a PV system of home $i$ in time slot $t$ ($1 \leq t \leq T$) in kWh can be measured as 
\begin{equation}
E_{i,RES}(t)= GHI_i(t) \cdot S_i \cdot \eta^{RES} \cdot \Delta t,
\label{RES_power}
\end{equation}
where $S_i$ is the total area $(m^2)$ of the solar panels of home $i$. $GHI_i(t)$ is the global horizontal irradiation $(kW/m^2)$ at the location of the solar panels of home $i$ in time slot $t$. $\eta^{RES}$ is the solar conversion efficiency of the PV system.

As illustrated in Fig. \ref{electricity_flow}, this energy can be used for appliances, ESS charging, and selling to the community. Thus, we determine the following constraint:
\begin{equation}
E_{i,RES}^{load}(t) + E_{i,RES}^{charge}(t) + E_{i,RES}^{sell}(t) = E_{i,RES}(t),
\label{RES_elements}
\end{equation}
where $E_{i,RES}^{load}(t) \geq 0 $ is the energy from the RES of home $i$ that is used for the appliances of this home in time slot $t$, and $E_{i,RES}^{sell}(t) \geq 0$ is the energy from the RES of home $i$ that is sold back to the community in time slot $t$.

\subsection{Energy Balancing}
To maintain the energy balance in a smart home, the total energy demand should be equal to the total energy supply. Hence, as shown in Fig. \ref{electricity_flow}, with $\forall t, \hspace{0.1cm}1 \leq t \leq T$, we obtain
\begin{equation}
E_{i,Fix}(t) + E_{i,HVAC}(t) = E_{i,Com}^{load}(t) + E_{i,ESS}^{load}(t) + E_{i,RES}^{load}(t),
\label{energy_balance}
\end{equation}
where $E_{i,Fix}(t) \geq 0$ is the fixed load of home $i$ in time slot $t$. Moreover, $E_{i,Com}^{load}(t) \geq 0$ is the energy that home $i$ requires for the home load from the community in time slot $t$.

\subsection{Avoiding Simultaneous Buying and Selling of Home}
In time slot $t$, let $E_{i,Com}^{buy}(t) \geq 0$ and $E_{i,Com}^{sell}(t) \geq 0$ be the total buying energy and selling energy of a home from/to the community, respectively. Thus, we obtain
\begin{equation}
E_{i,Com}^{buy}(t) = E_{i,Com}^{load}(t) + E_{i,Com}^{charge}(t) 
\end{equation}
\begin{equation}
E_{i,Com}^{sell}(t) = E_{i,RES}^{sell}(t) + E_{i,ESS}^{sell}(t). 
\end{equation}

Residents are unable to buy and sell from/to the community simultaneously within a time slot. Hence, we determine the following constraints:
\begin{equation}
0 \leq E_{i,Com}^{buy}(t) \leq M \cdot mode_{i,home}(t)
\label{constraint_buy_each_home}
\end{equation}
\begin{equation}
0 \leq E_{i,Com}^{sell}(t) \leq M \cdot \big(1 - mode_{i,home}(t)\big)
\label{constraint_sell_each_home}
\end{equation}
\begin{equation}
mode_{i,home}(t) = 
   \begin{cases}
   1 & \text{if home i buys energy from community } \\
   0 & \text{if home i sells energy to community }, \\
   \end{cases}
\label{buy_sell_mode}
\end{equation}
where $M$ is a very large value that $E_{i,Com}^{buy}(t)$ and $E_{i,Com}^{sell}(t)$ never exceed; e.g., $M=10^9$. $mode_{i,home}(t)$ is a binary variable to avoid the simultaneous buying and selling of home $i$ in time slot $t$.

\subsection{Optimization Problem of Community}
In time slot $t$, the buying/selling energy of home $E_{i,Com}(t)$ from/to the community is 
\begin{equation}
E_{i,Com}(t) =  E_{i,Com}^{buy}(t) - E_{i,Com}^{sell}(t).
\end{equation}

The total buying/selling energy of the community from/to the EP in time slot $t$ is
\begin{align}
E_{Com}(t) = \displaystyle \sum_{i=1}^{N}E_{i,Com}(t) = \displaystyle \sum_{i=1}^{N}E_{i,Com}^{buy}(t) - \displaystyle \sum_{i=1}^{N}E_{i,Com}^{sell}(t).
\end{align}

If $E_{Com}(t) > 0$, the total buying energy is larger than the total selling energy within the community and it needs to buy an energy quantity $E_{Com}(t)$ from the EP. If $E_{Com}(t) < 0$, the total buying energy is smaller than the total selling energy within the community and it needs to sell an energy quantity $|E_{Com}(t)|$ to the EP. If $E_{Com}(t) = 0$, the total buying energy is equal to the total selling energy within the community.

Let $P_{MG}(t)$ be the day-ahead buying price at which the community buys electricity from the EP in time slot $t$, and $\alpha \cdot P_{MG}(t)$ with $0 < \alpha < 1$ be the day-ahead selling price at which the community sells to the EP in time slot $t$. The optimization problem of the community that minimizes the daily energy cost of the community can be formulated as
\begin{align}
&min \displaystyle \sum_{t=1}^{T} P(t) \cdot E_{Com}(t) \nonumber \\
&= min \displaystyle \sum_{t=1}^{T}P(t) \cdot\Big(\displaystyle \sum_{i=1}^{N}E_{i,Com}^{buy}(t) - \displaystyle \sum_{i=1}^{N}E_{i,Com}^{sell}(t)\Big),
\label{optimiazation_problem_1}
\end{align}
where
\begin{equation}
P(t) = \begin{cases}
    P_{MG}(t) & \quad \text{if } \displaystyle \sum_{i=1}^{N}E_{i,Com}^{buy}(t) > \displaystyle \sum_{i=1}^{N}E_{i,Com}^{sell}(t) \\
    \alpha \cdot P_{MG}(t)  & \quad \text{if } \displaystyle \sum_{i=1}^{N}E_{i,Com}^{buy}(t) < \displaystyle \sum_{i=1}^{N}E_{i,Com}^{sell}(t) 
  \end{cases}
\label{price_definition}
\end{equation}
subject to:
\begin{equation}
(\ref{constraint_power_HVAC}) - (\ref{buy_sell_mode})
\end{equation}
\begin{equation}
-E_{max} \leq\displaystyle \sum_{i=1}^{N}E_{i,Com}^{buy}(t) - \displaystyle \sum_{i=1}^{N}E_{i,Com}^{sell}(t) \leq E_{max},
\label{thresold}
\end{equation}
where constraint (\ref{thresold}) is added to restrict the buying/selling energy of the community to be lower than an energy peak $E_{max}$ (the capacity of the power cable).

\subsection{Linear Transformation}
Owing to the definition of variable $P(t)$ in (\ref{price_definition}), our optimization problem in (\ref{optimiazation_problem_1}) is a nonlinear function that can be solved using numerous well-known nonlinear solvers and heuristic algorithms. However, these usually require substantial computational time, especially for a large community including many homes. Hence, in this section, we describe a means of transforming our optimization problem into an MILP problem.

To remove variable $P(t)$, we introduce a new variable $s(t)$, which is a binary variable that indicates the buying/selling status of the community from/to the EP at each time slot $t$.
\begin{equation}
s(t) = \begin{cases}
    1 & \quad \text{if } \displaystyle \sum_{i=1}^{N}E_{i,Com}^{buy}(t) > \displaystyle \sum_{i=1}^{N}E_{i,Com}^{sell}(t) \\
    0  & \quad \text{if } \displaystyle \sum_{i=1}^{N}E_{i,Com}^{buy}(t) < \displaystyle \sum_{i=1}^{N}E_{i,Com}^{sell}(t) 
  \end{cases}
\label{buy_or_sell}
\end{equation}

We obtain the following constraints for variable $s(t)$:
\begin{equation}
\displaystyle \sum_{i=1}^{N}E_{i,Com}^{buy}(t) \geq \displaystyle \sum_{i=1}^{N}E_{i,Com}^{sell}(t) - M \cdot (1-s(t))
\label{buy_or_sell_1}
\end{equation}
\begin{equation}
\displaystyle \sum_{i=1}^{N}E_{i,Com}^{buy}(t) \leq \displaystyle \sum_{i=1}^{N}E_{i,Com}^{sell}(t) + M \cdot s(t),
\label{buy_or_sell_2}
\end{equation}
where $M$ is a very large value that $\displaystyle \sum_{i=1}^{N}E_{i,Com}^{buy}(t)$ and $\displaystyle \sum_{i=1}^{N}E_{i,Com}^{sell}(t)$ never exceed; e.g., $M=10^9$.

Our optimization problem can be transformed into
\begin{equation}
min \displaystyle \sum_{t=1}^{T} C_{Com}(t),
\label{optimiazation_problem_2}
\end{equation}
where variable $C_{Com}(t)$ is the energy cost of the community in time slot $t$ and it has the following constraints:
\begin{equation}
C_{Com}(t) \geq (\displaystyle \sum_{i=1}^{N}E_{i,Com}^{buy}(t) - \displaystyle \sum_{i=1}^{N}E_{i,Com}^{sell}(t)) \cdot P_{MG}(t) - M \cdot (1-s(t))
\label{buy_or_sell_3}
\end{equation}
\begin{equation}
C_{Com}(t) \leq (\displaystyle \sum_{i=1}^{N}E_{i,Com}^{buy}(t) - \displaystyle \sum_{i=1}^{N}E_{i,Com}^{sell}(t)) \cdot P_{MG}(t) + M \cdot (1-s(t))
\label{buy_or_sell_4}
\end{equation}
\begin{equation}
C_{Com}(t) \geq (\displaystyle \sum_{i=1}^{N}E_{i,Com}^{buy}(t) - \displaystyle \sum_{i=1}^{N}E_{i,Com}^{sell}(t)) \cdot \alpha \cdot P_{MG}(t) - M \cdot s(t)
\label{buy_or_sell_5}
\end{equation}
\begin{equation}
C_{Com}(t) \leq (\displaystyle \sum_{i=1}^{N}E_{i,Com}^{buy}(t) - \displaystyle \sum_{i=1}^{N}E_{i,Com}^{sell}(t)) \cdot \alpha \cdot P_{MG}(t) + M \cdot s(t)
\label{buy_or_sell_6}
\end{equation}
subject to:
\begin{equation}
(\ref{constraint_power_HVAC}) - (\ref{buy_sell_mode}), (\ref{thresold}), (\ref{buy_or_sell_1}), (\ref{buy_or_sell_2}), (\ref{buy_or_sell_3}) - (\ref{buy_or_sell_6}).
\end{equation}

It is clear that problem (\ref{optimiazation_problem_2}) is an MILP problem that can be solved by many advanced mathematical solvers in a short time. The output of solving the problem is the day-ahead schedule of all DERs and HVACs in the community to minimize the daily energy cost of the community and satisfy the thermal comfort in every home.

\section{Local Energy Trading Mechanism}
In this section, we introduce local energy trading among the homes of the community. This trading function is run after the OS is run, and each home already knows the buying or selling action at each time slot of the day. As illustrated in Fig. \ref{community}, the residents of the homes first trade energy with others in the community using the local buying and selling prices, instead of trading directly with the EP. After trading together within the community, if the community requires more energy or has surplus energy to sell, the community trades directly with the EP. To encourage energy trading among the homes of the community, the local buying/selling prices that are proposed by the local trading manager of the central operation unit should be smaller/larger than the buying/selling prices, namely $P_{MG}(t)$/$\alpha \cdot P_{MG}(t)$, that are proposed by the EP. The pricing in this study is the MMR pricing, which is adopted from \cite{ye2021scalable}, \cite{li2020computationally}.

\subsection{MMR Pricing}
Let $P_{LB}(t), P_{LS}(t)$ be the local buying and selling prices at time slot $t$, respectively, and $P_{mid}(t)$ be a price that has a value in the range from the selling to buying price proposed by the EP.
\begin{equation}
\alpha \cdot P_{MG}(t) \leq P_{mid}(t) \leq P_{MG}(t)
\end{equation}

In this study, we use $P_{mid}(t)= (1+\alpha)\cdot P_{MG}(t)/2$. At each time slot $t$, the local buying and selling prices depend on the variable $E_{Com}(t)$ of the community, and they can be calculated as follows.
\begin{enumerate}
\item $E_{Com}(t)=0$: Within the community, the total buying energy is equal to the total selling energy. The local buying and selling prices at this time slot are equal to the average price $P_{mid}(t)$.
\begin{equation}
P_{LB}(t) = P_{LS}(t) = P_{mid}(t)
\end{equation}
\item $E_{Com}(t) > 0$: Within the community, the total buying energy is larger than the total selling energy. The community needs to buy an energy quantity $E_{Com}(t)$ from the EP at buying price $P_{MG}(t)$. In this case, the local selling and buying prices in this time slot will be
\begin{equation}
P_{LS}(t) = P_{mid}(t)
\end{equation}
\begin{equation}
P_{LB}(t) = \frac{P_{LS}(t) \displaystyle \sum_{i=1}^{N}E_{i,Com}^{sell}(t) + P_{MG}(t) E_{Com}(t)}{\displaystyle \sum_{i=1}^{N}E_{i,Com}^{buy}(t)}.
\label{local_buying_price}
\end{equation}

Formula (\ref{local_buying_price}) indicates that the total buying payment is proportionally shared among the buying homes according to their buying energy quantity.
\item $E_{Com}(t) < 0$: Within the community, the total buying energy is smaller than the total selling energy. The community needs to sell an energy quantity $|E_{Com}(t)|$ to the EP at selling price $\alpha \cdot P_{MG}(t)$. In this case, the local buying and selling prices in this time slot will be
\begin{equation}
P_{LB}(t) = P_{mid}(t)
\end{equation}
\begin{equation}
P_{LS}(t) = \frac{P_{LB}(t) \displaystyle \sum_{i=1}^{N}E_{i,Com}^{buy}(t) + \alpha P_{MG}(t) |E_{Com}(t)|}{\displaystyle \sum_{i=1}^{N}E_{i,Com}^{sell}(t)}.
\label{local_selling_price}
\end{equation}

Formula (\ref{local_selling_price}) indicates that the total selling profit is proportionally shared among the selling homes according to their selling energy quantity.
\end{enumerate}

With this pricing scheme, it is clear that the local buying and selling prices within our community are always better than the buying and selling prices that are proposed by the EP. Hence, to improve the benefit for each home, the central operation unit prefers the trading of energy among its homes over trading with the EP. Given the local buying and selling prices, the daily energy cost of home $i$ can be calculated as follows:
\begin{align}
C_{i}= \displaystyle \sum_{t=1}^{T} P_{L}(t) \cdot E_{i,Com}(t),
\end{align}
where
\begin{equation}
P_{L}(t) = \begin{cases}
    P_{LB}(t) & \quad \text{if } E_{i,Com}(t) > 0 \\
    P_{LS}(t)  & \quad \text{if } E_{i,Com}(t) < 0. 
  \end{cases}
\end{equation}

\section{Prosumer-centric CEMS for Community}
We describe a prosumer-centric CEMS for a community for comparison with our proposed CEMS in this section. As discussed in Section I, in a prosumer-centric CEMS, each prosumer is first allowed to optimize their own optimization problem without energy trading. After solving this problem, a day-ahead schedule of their individual DERs is provided, following which they join a local energy market to receive further benefits. 

In the first stage, the optimization problem of each home in the community is as follows:
\begin{align}
&min \displaystyle \sum_{t=1}^{T} P_{i}(t) \cdot E_{i,Com}(t) \nonumber \\
&= min \displaystyle \sum_{t=1}^{T}P_{i}(t) \cdot\Big(E_{i,Com}^{buy}(t) - E_{i,Com}^{sell}(t)\Big),
\label{optimiazation_problem_each_home}
\end{align}
where
\begin{equation}
P_{i}(t) = \begin{cases}
    P_{MG}(t) & \quad \text{if }  E_{i,Com}(t) >0 \\
    \alpha \cdot P_{MG}(t)  & \quad \text{if } E_{i,Com}(t) < 0 
  \end{cases}
\label{price_definition_each_home}
\end{equation}
subject to:
\begin{equation}
(\ref{constraint_power_HVAC}) - (\ref{buy_sell_mode})
\end{equation}
\begin{equation}
-E_{i,max} \leq E_{i,Com}^{buy}(t) - E_{i,Com}^{sell}(t) \leq E_{i,max},
\label{thresold_each_home}
\end{equation}
in which constraint (\ref{thresold_each_home}) is added to restrict the buying/selling energy of the home to be lower than the energy peak $E_{i,max}$.

As a home cannot simultaneously buy and sell energy in a time slot (constraints (\ref{constraint_buy_each_home}) and (\ref{constraint_sell_each_home})), the problem in (\ref{optimiazation_problem_each_home}) can be converted into
\begin{equation}
min \displaystyle \sum_{t=1}^{T}\Big(E_{i,Com}^{buy}(t) \cdot P_{MG}(t) - E_{i,Com}^{sell}(t) \cdot \alpha \cdot P_{MG}(t) \Big)
\label{optimiazation_problem_each_home_2}
\end{equation}
subject to:
\begin{equation}
(\ref{constraint_power_HVAC}) - (\ref{buy_sell_mode}), (\ref{thresold_each_home}).
\end{equation}

It is clear that the problem in (\ref{optimiazation_problem_each_home_2}) is an MILP problem that can be solved by many solvers. The output is the day-ahead schedule of the DERs and HVAC of the home.

In the second stage, based on this day-ahead schedule, each prosumer joins a local energy market. For comparison, the local energy market with the MMR pricing described in Section IV is reused in this CEMS.

\section{Simulations and Results}

In this section, a community consisting of a group of $10$ homes is simulated under the control of different CEMSs: the proposed CEMS, a prosumer-centric CEMS, and no CEMS in which each home trades directly with the EP, to verify the efficiency and the performance of our proposed CEMS.

\subsection{Simulation Setup}
\begin{figure}[h]
\centering
\includegraphics[scale=1]{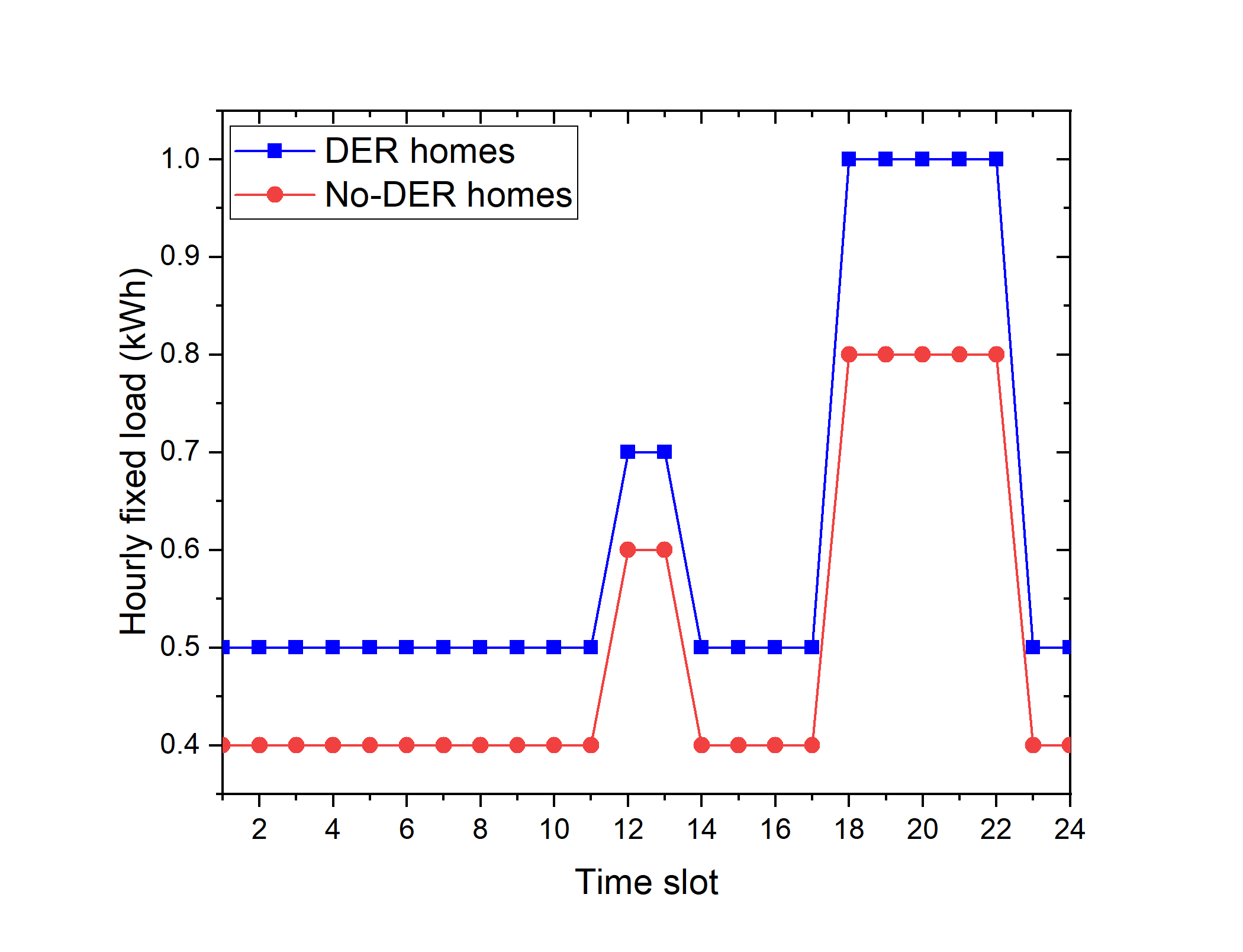}
\caption{Hourly fixed loads of DER homes and no-DER homes.}
\label{load_homes}
\end{figure}

Our proposed community consists of 10 homes in Detroit city, Michigan, USA. Each home has its own HVAC as the controllable load and the parameters of the HVAC are the same for every home, as indicated in Table \ref{HVAC_parameters}. For the ESS and RES, we assume that three homes have an ESS and a PV system: \textit{Home 1, Home 2, and Home 3}; two homes only have a PV system: \textit{Home 4 and Home 5}; two homes only have an ESS: \textit{Home 6 and Home 7}; and three homes have neither: \textit{Home 8, Home 9 and Home 10}, as listed in Table \ref{spec_homes}. The numbers in this Table indicate the area of the solar panels $S_i$ and the maximum level of the ESS $EL_{i,max}$ at each home. The initial level $EL_{i,0}$ and minimum level $EL_{i,min}$ of the ESS at each home are the same, and $EL_{i,min} = EL_{i,0}=0.5$ kWh. The maximum charging rate $Ch_{i,rate}$ and maximum discharging rate $Dh_{i,rate}$ of the ESS at each home are the same, and $Ch_{i,rate}=Dh_{i,rate}=2$ kW/h. $\eta^{ESS} = \eta^{RES}=0.9$. The hourly fixed loads of the DER homes that have at least one DER (\textit{Home 1 to Home 7}) and the homes without DERs (\textit{Home 8 to Home 10}) used in our simulation are shown in Fig. \ref{load_homes}. It is worth noting that our simulations can be run well with any values of hourly fixed loads of homes in the real life.
\begin{table*}[h]
\caption{PV system and ESS at each home.}
\begin{center}
\begin{tabular}{|c|c|c|c|c|c|c|c|c|c|c|}
\hline
\textbf{Device} &\textbf{Home 1} &\textbf{Home 2} &\textbf{Home 3} &\textbf{Home 4} &\textbf{Home 5} &\textbf{Home 6} &\textbf{Home 7} &\textbf{Home 8} &\textbf{Home 9} &\textbf{Home 10}\\
\hline
PV (\textit{$m^2$}) & 10 & 10 & 10 & 5 & 5 & X & X & X & X & X \\
\hline
ESS (\textit{kWh}) & 10 & 10 & 10 & X  & X  & 5 & 5 & X & X & X \\
\hline
\end{tabular}
\label{spec_homes}
\end{center}
\end{table*}

\begin{table}[h]
\caption{Main parameters of HVAC system and environment.}
\centering
\begin{tabular}{|c|c|}
\hline
\textbf{Parameter}	& \textbf{Value}\\
\hline
$T_{min}$ &  $66.2$ $^\text{o}$F ($19$ $^\text{o}$C)\\
\hline
$T_{max}$ &  $75.2$ $^\text{o}$F ($24^\text{o}$C)\\
\hline
$P_{max}$ &  $15$ kW\\
\hline
$\varepsilon$ &  $0.7$ \cite{yu2019deep}\\
\hline
$\eta$ &  $2.5$ \cite{yu2019deep}\\
\hline
$A$ &  $0.14$ kW/$^\text{o}$F \cite{yu2019deep}\\
\hline

\end{tabular}
\label{HVAC_parameters}
\end{table} 

\begin{figure}[]
\centering
\includegraphics[scale=1]{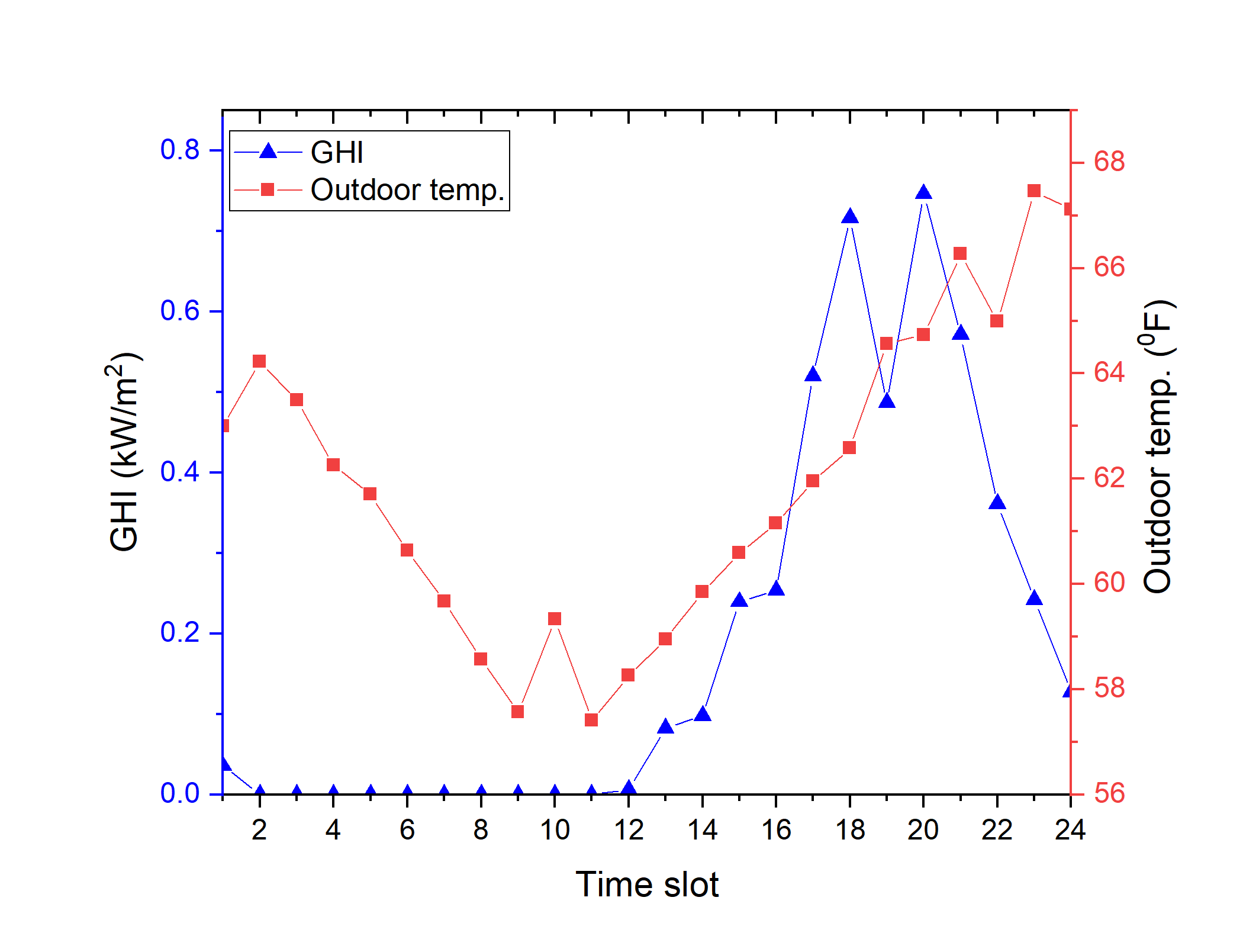}
\caption{Day-ahead GHI and outside temperature of day in Detroit city.}
\label{GHI}
\end{figure}

\begin{figure}[]
\centering
\includegraphics[scale=1]{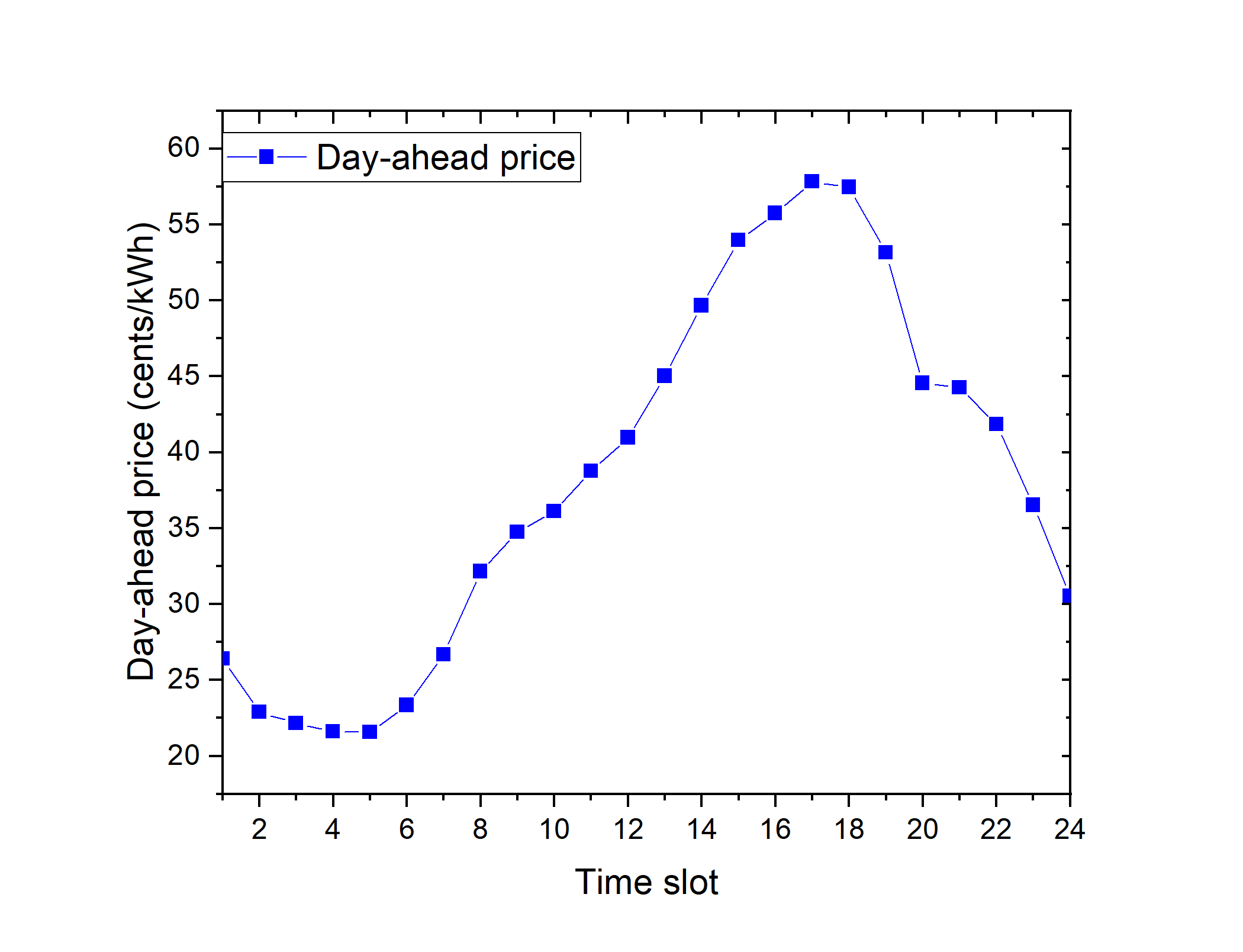}
\caption{Day-ahead prices of day in Michigan state.}
\label{DAP}
\end{figure}

The day-ahead GHI from a photovoltaic geographical information system database \cite{irradiation} and day-ahead outside temperature from the Kaggle website \cite{kaggle} of a day in Detroit city, Michigan state, USA are presented in Fig. \ref{GHI}. The day-ahead hourly prices $P_{MG}(t)$ of the EP in Detroit city, which were extracted from the Pecan Street database \cite{prices}, are also indicated in Fig. \ref{DAP}. We assume that the day-ahead selling price from the community to the EP is equal to $80\%$ of this price at every time slot ($\alpha = 0.8$).

All of our simulations for the three scenarios were built and run using the mathematical programming software AIMMS $v4.82$ \cite{AIMMS} with CPLEX $v20.1$ installed on an Intel(R) Core(TM) i7-8700 CPU @ 3.20GHz and 16 GB RAM with Windows $10$ Pro (64-bit). AIMMS is a high-level modeling commercial software that integrates advanced mathematical solvers such as CPLEX and Conopt for solving LP, MILP, and MINLP problems.

\subsection{Energy Cost of Community with Three CEMSs}
\begin{figure}[]
\centering
\includegraphics[scale=1]{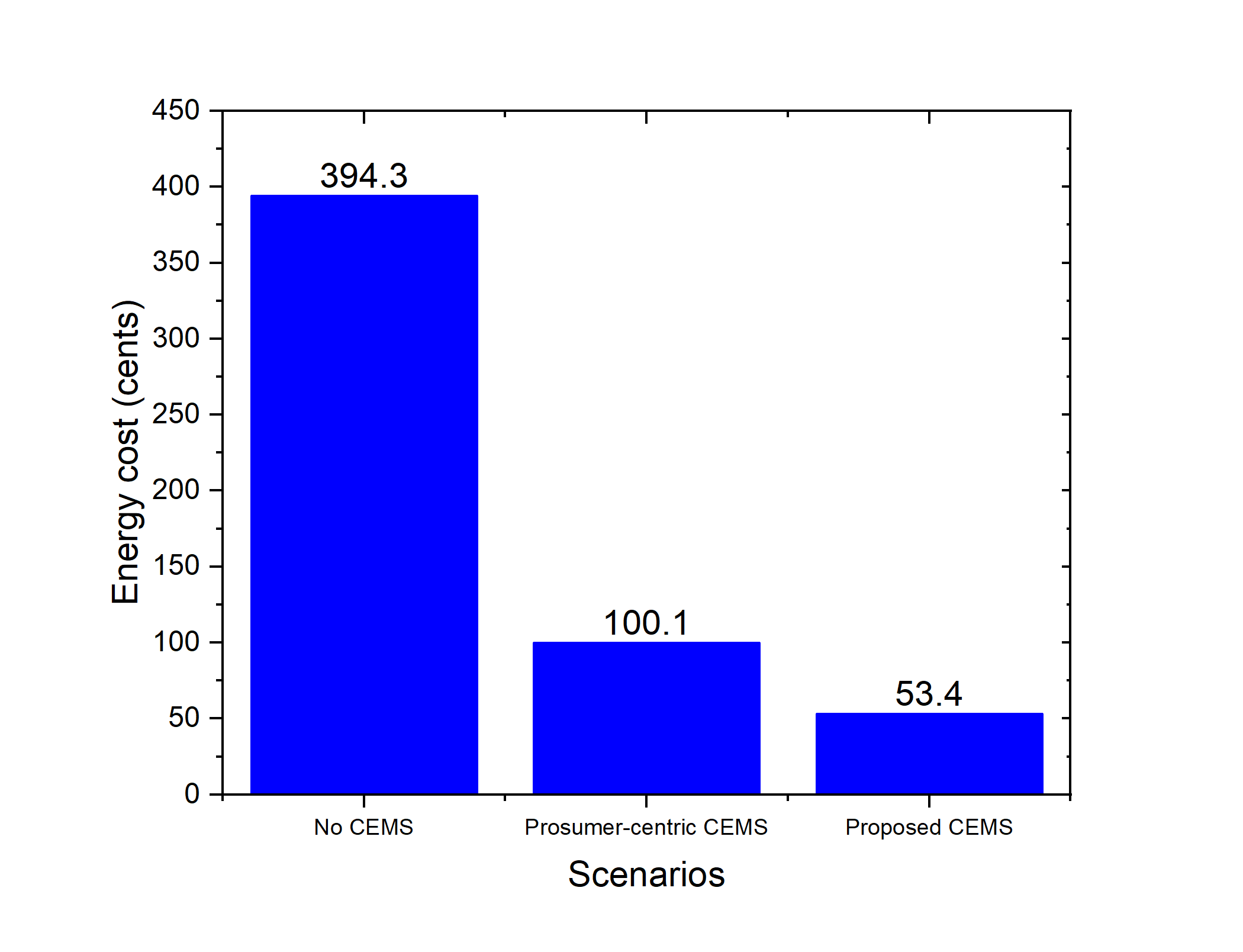}
\caption{Daily energy cost of community with three CEMSs.}
\label{energy_cost}
\end{figure}

\begin{figure}[]
\centering
\includegraphics[scale=1]{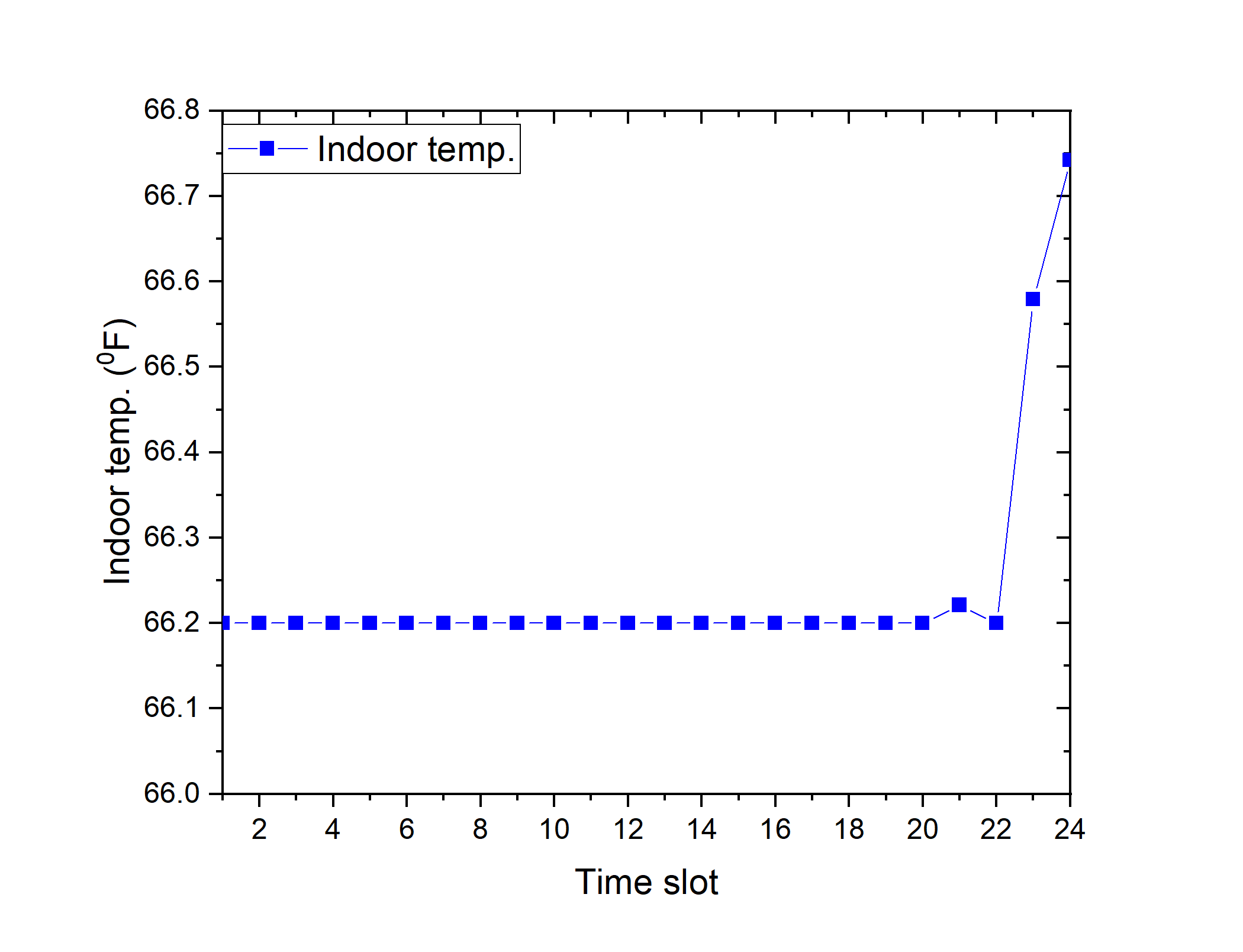}
\caption{Indoor temperature during day at every home of community.}
\label{indoor_temp}
\end{figure}

Fig. \ref{energy_cost} depicts the energy cost of our proposed community with three CEMSs for a day, whereas the indoor temperature during this day, which is the same at every home, is illustrated in Fig. \ref{indoor_temp}. As indicated in these figures, the thermal comfort is satisfied at every home of the community; however, the energy cost of the community with no CEMS and the prosumer-centric CEMS are $394.3$ cents and $100.1$ cents, respectively, whereas the energy cost of the community with the proposed CEMS is only $53.4$ cents, which is a significant decrease of $50\%$ compared to that of the community with the prosumer-centric CEMS. The community with the proposed CEMS is the best among the three scenarios in terms of the daily energy cost.

\begin{figure}[]
\centering
\includegraphics[scale=1]{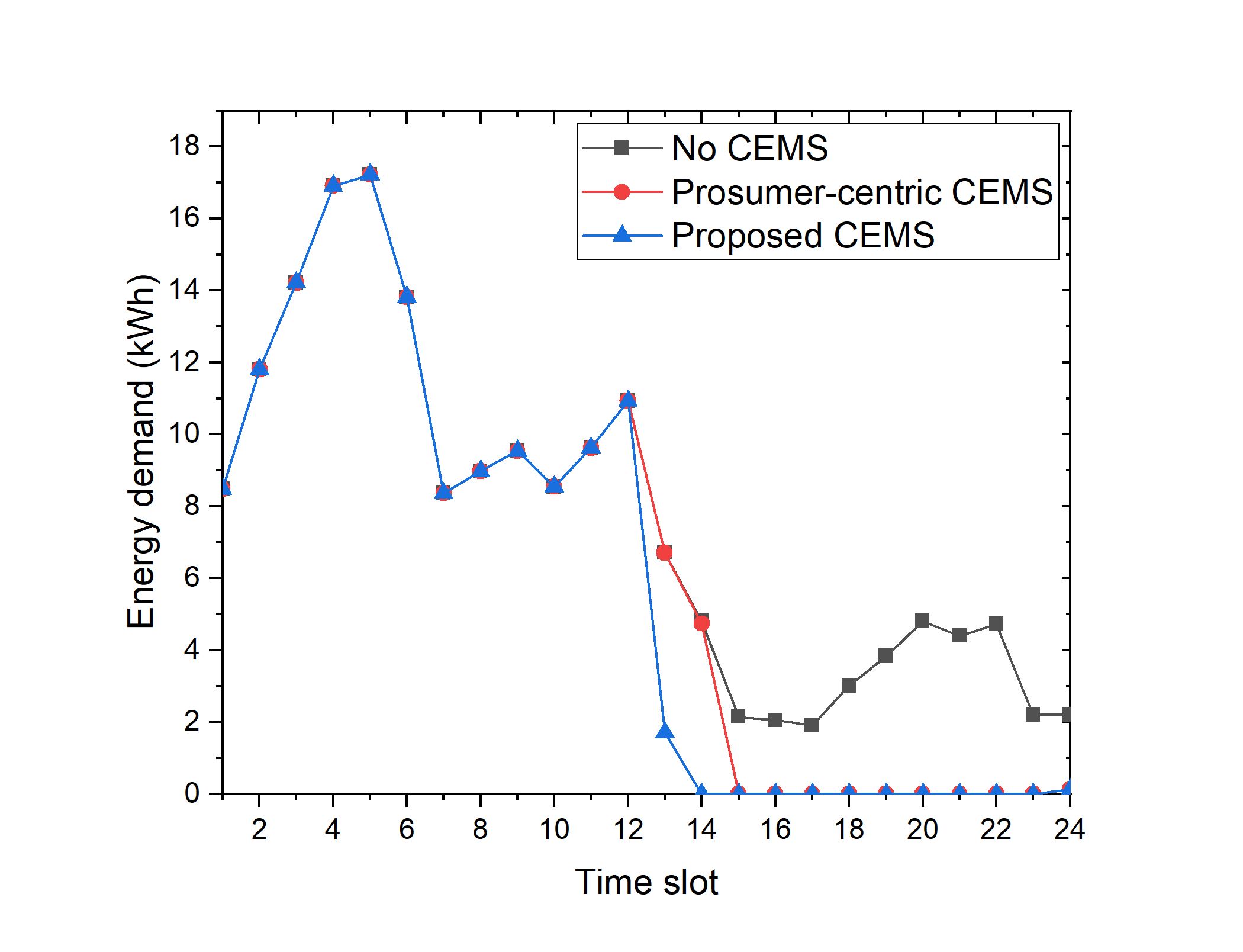}
\caption{Energy demand of community with three CEMSs from EP.}
\label{energy_demand}
\end{figure}

\begin{figure}[]
\centering
\includegraphics[scale=1]{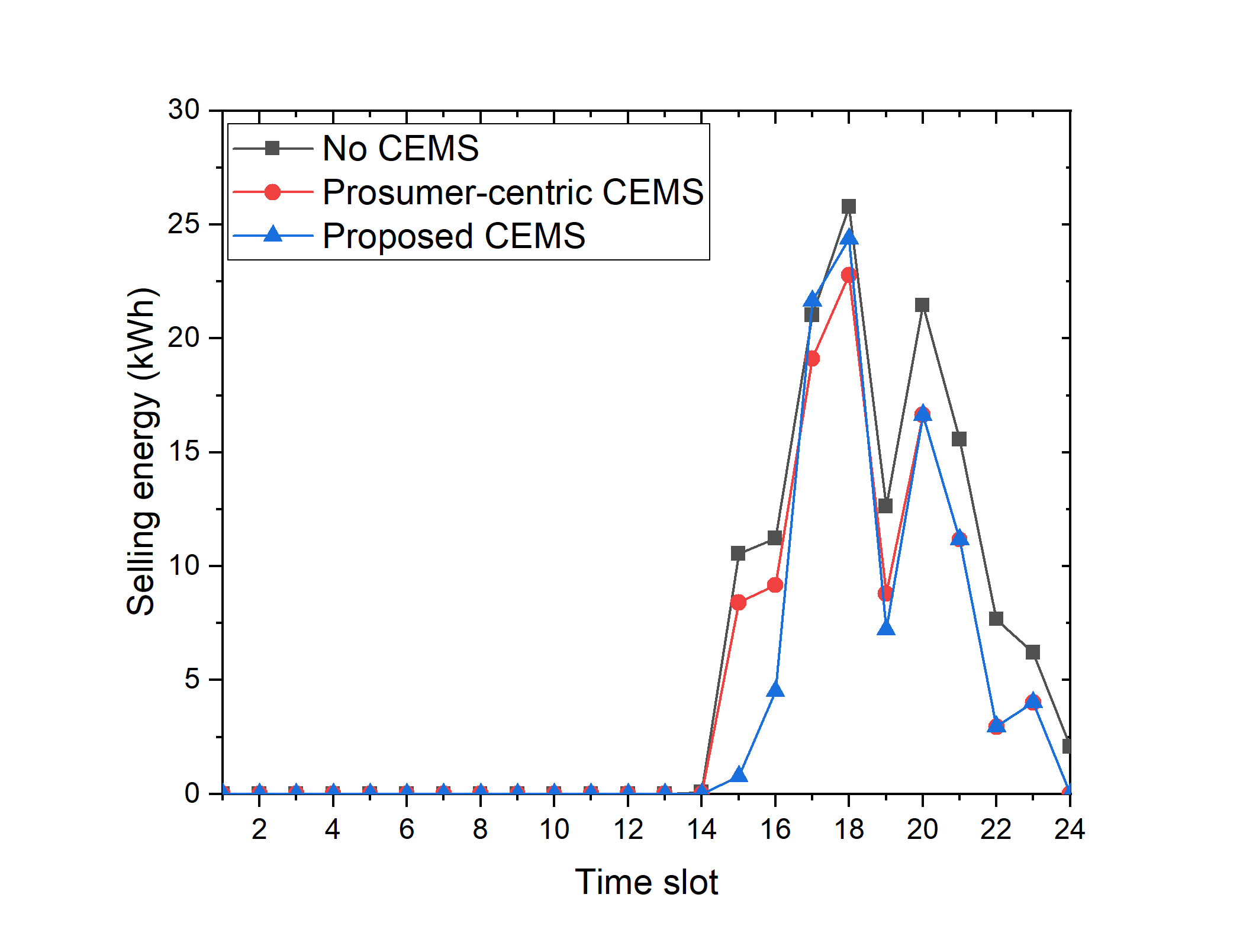}
\caption{Selling energy of community with three CEMSs to EP.}
\label{selling_energy}
\end{figure}

The energy demand and selling energy of the community with the three CEMSs from/to the EP at each time slot are shown in Figs. \ref{energy_demand} and \ref{selling_energy}, respectively, to gain better insight into the operation of the community. As illustrated in Fig. \ref{energy_demand}, from time slot $1$ to time slot $12$, the energy demand of the community with three CEMSs from the EP is almost the same because the GHI at these time slots is almost $0$. However, following time slot $12$, there are differences in the operations of the three scenarios when the GHI at these time slots is available. In the no-CEMS community, certain homes still require energy from the EP because they cannot buy it from other homes where the surplus energy has to be sold back to the EP. In the community with the prosumer-centric CEMS and proposed CEMS, from time slot $15$ to time slot $24$, the homes do not require energy from the EP because they can share energy when they have surplus energy from their PV systems or ESSs. However, at time slots $13$ and $14$, the energy demand of the community with the proposed CEMS is smaller than that of the community with the prosumer-centric CEMS. Thus, the daily energy cost of the community with the proposed CEMS smaller than that of the community with the prosumer-centric CEMS.

Fig. \ref{selling_energy} depicts the selling energy of the community with three CEMSs. This result is another reason that the daily energy cost of the community with the proposed CEMS is smaller than that of the others. Starting from time slot $15$, the community with both the prosumer-centric CEMS and proposed CEMS sells its surplus energy to the EP. However, the community with the proposed CEMS sells more energy than the community with the prosumer-centric CEMS at high-price time slots (e.g., time slots $17$ and $18$). In the no-CEMS scenario, the selling energy is greater than that of two cases in these times slots. However, the energy demand of the community in the no-CEMS scenarios is also substantially larger than that of the community in the other scenarios at these time slots, as indicated in Fig. \ref{energy_demand}, and the selling price is always lower than the buying price at every time slot. Hence, the daily energy cost of the community in the no-CEMS scenario is the highest among the three scenarios.

\subsection{Computational Time of Community with Proposed CEMS}
Table \ref{computational_time} displays the computational time of the optimization problem of the community with the proposed CEMS in four cases: 10 homes, 50 homes, 100 homes, and 500 homes.

\begin{table}[ht]
\caption{Computational time of community with proposed CEMS.}
\centering
\begin{tabular}{|c|c|c|c|c|}
\hline
\textbf{Number of homes} &10 homes	& 50 homes &100 homes &500 homes\\
\hline
\textbf{Time(s)} &1 & 3.3 & 10.8 & 118.2\\
\hline
\end{tabular}
\label{computational_time}
\end{table} 

Our optimization problem is an MILP problem, which is generally a type of NP-hard problem. To the best of our knowledge, no polynomial-time algorithm is available that can be applied to solve all MILP problems. However, not every MILP is an NP-hard problem. Moreover, at present, efficient algorithms are available in which certain MILP problems can be relaxed and solved within a reasonable amount of time when they are combined with advanced mathematical solvers (e.g., CPLEX) in commercial software. In our optimization problem, the computational times for the 100-home and 500-home communities are $10.8$ s and $118.2$ s, respectively, in the AIMMS software. It is clear that these times are sufficiently small for building a day-ahead schedule for a community.

\subsection{Daily Energy Cost of Each Home in Community with Three CEMSs}
\begin{figure}[]
\centering
\includegraphics[scale=1]{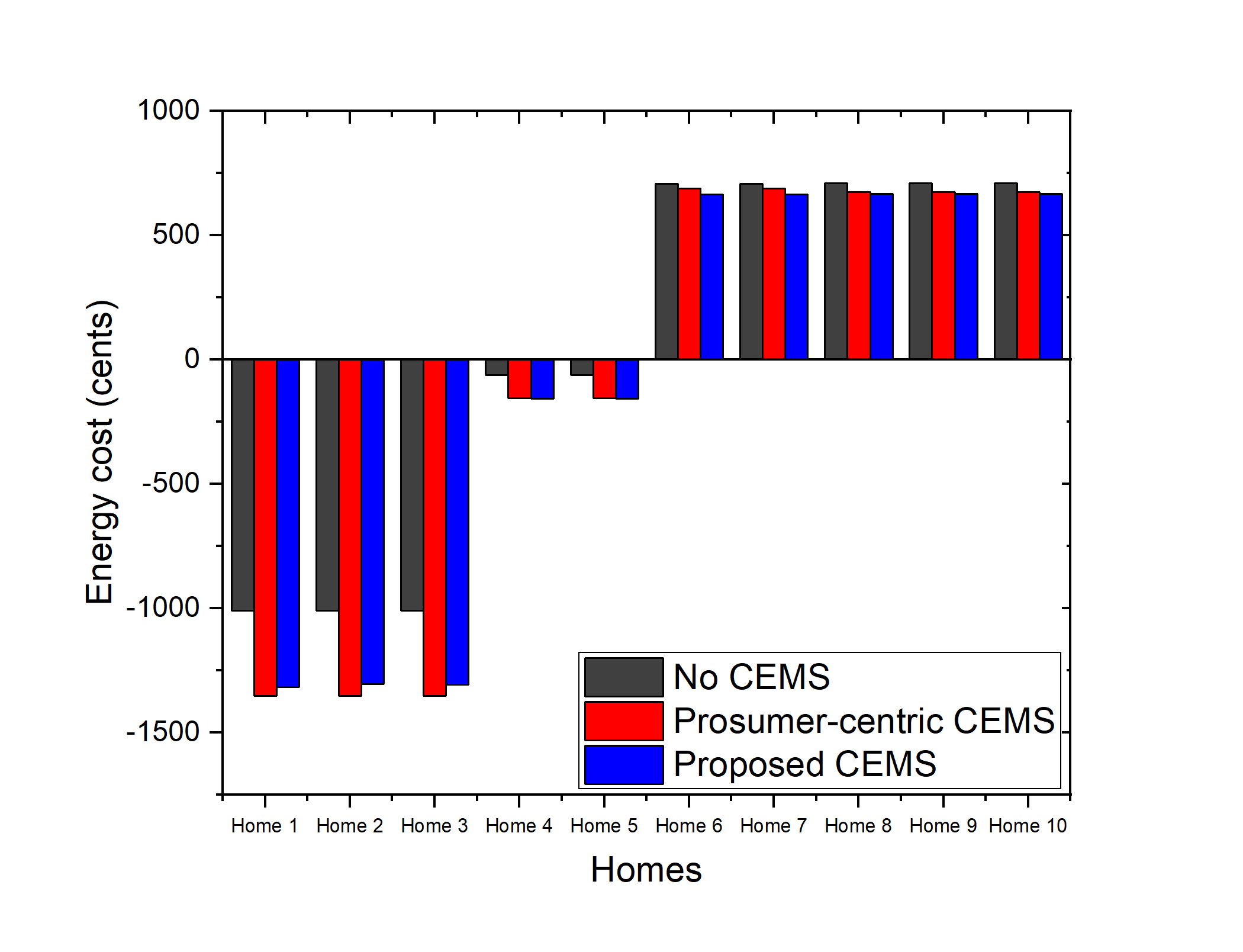}
\caption{Energy cost of each home in community with three CEMSs.}
\label{cost_home}
\end{figure}

The daily energy cost of each home in the community with three CEMSs is shown in Fig. \ref{cost_home}. In this figure, the daily energy costs of the homes that have PV systems are the negative values. This means that residents of these homes receive some money because the energy that is generated by the PV system is greater than the energy demand of these homes. As the local buying/selling prices are better than the buying/selling prices proposed by the EP, this cost of each home in the community with the prosumer-centric CEMS and proposed CEMS is always better than that of each home in the community with no CEMS, where every home must directly trade its surplus energy with the EP. According to this figure, the daily benefits of the homes that have PV systems and ESSs (\textit{Home 1 to Home 3}) in the community with the proposed CEMS are worse than those in the community with the prosumer-centric CEMS; however, the daily benefits of the remaining homes in the community with the proposed CEMS are better than those in the community with the prosumer-centric CEMS. This is because, in the community with the proposed CEMS, the objective function minimizes the overall energy cost of the community. Hence, the homes with PV systems and ESSs must sacrifice their benefits and share their benefits with the other homes to achieve the minimum overall energy cost of the community. The OS of the central operation unit schedules the DERs at the homes with PV systems and ESSs by considering not only the loads at these homes, but also the loads at other homes in the community. In contrast, in the community with the prosumer-centric CEMS, the objective function minimizes the energy cost of each home and the schedule of the DERs at homes with PV systems and ESSs that is generated based only on the loads of this home.

\begin{figure}[]
\centering
\includegraphics[scale=0.35]{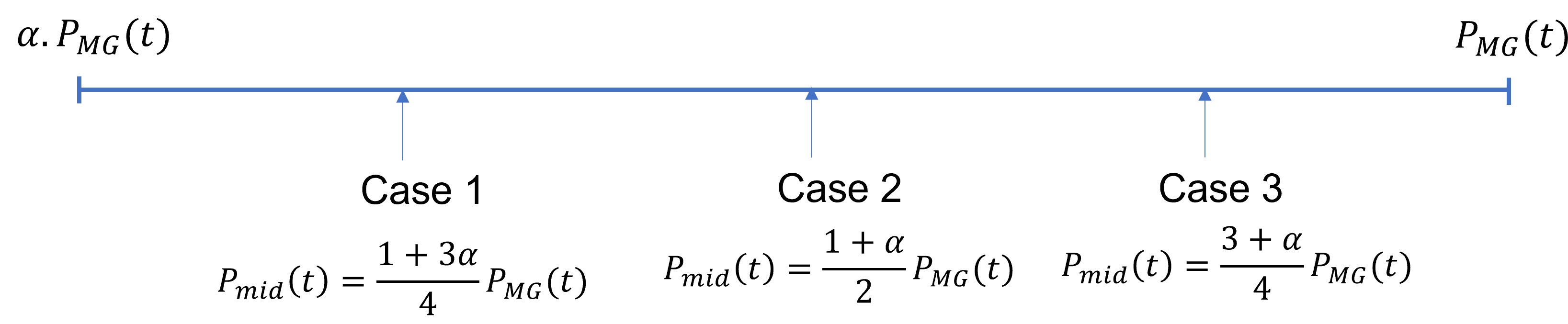}
\caption{Values of $P_{mid}(t)$ considered in three cases.}
\label{middle_prices}
\end{figure}

A special feature of MMR pricing is that the local buying/selling prices can be changed by the central operation unit, and the daily benefits of the homes in the community with the proposed CEMS can be flexibly increased or decreased. Using this feature, the central operation unit can improve the daily benefits of homes that have PV systems. We consider the daily energy cost of the homes in the community with the proposed CEMS under different values of variable $P_{mid}(t)$, as shown in Fig. \ref{middle_prices}.

\begin{equation}
P_{mid}(t) = \begin{cases}
    (1+3 \cdot \alpha)\cdot P_{MG}(t)/4 & \quad \text{Case 1} \\
    (1+\alpha)\cdot P_{MG}(t)/2  & \quad \text{Case 2} \\
    (3+\alpha)\cdot P_{MG}(t)/4 & \quad \text{Case 3}
  \end{cases}
\end{equation}

\begin{figure}[]
\centering
\includegraphics[scale=1]{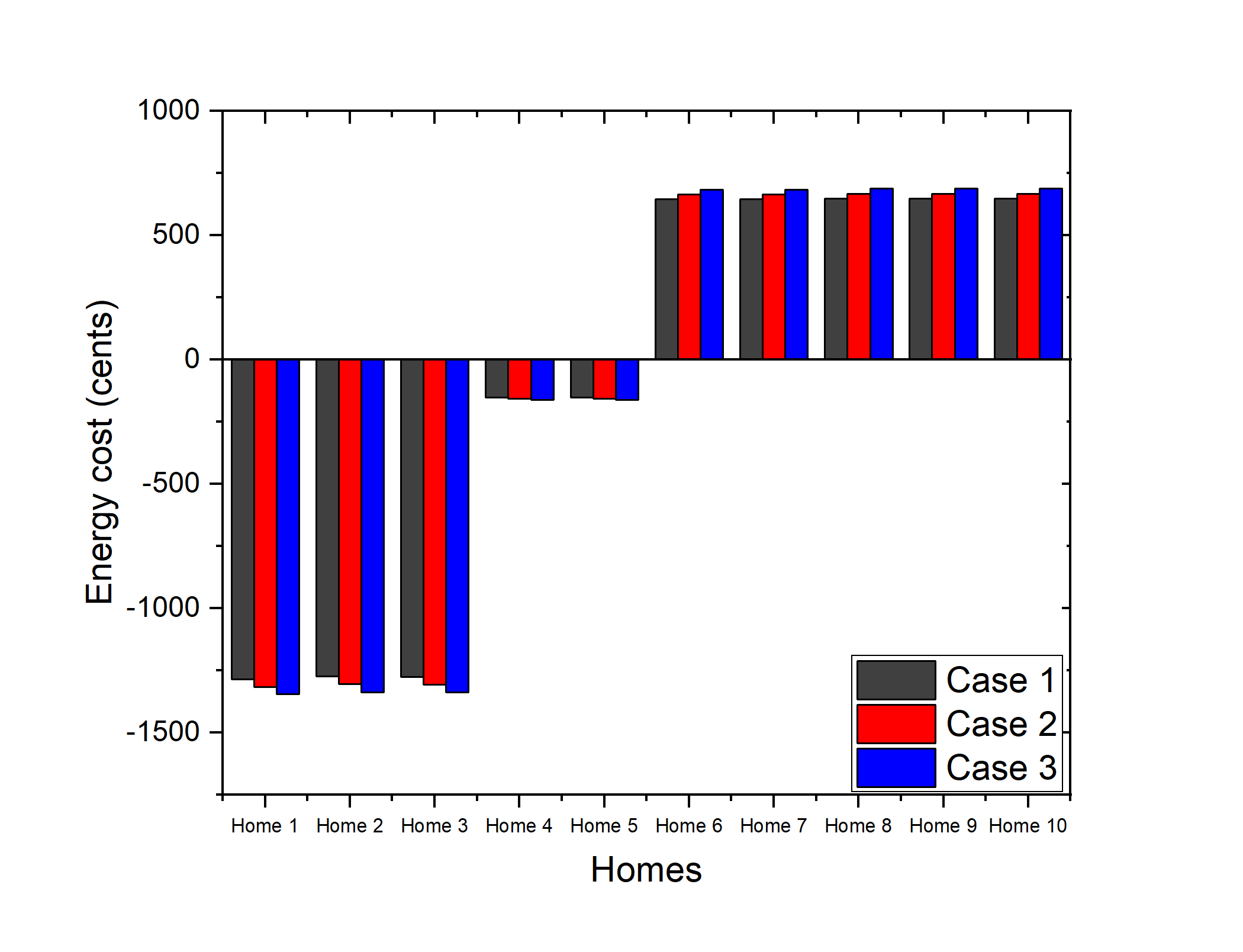}
\caption{Trade-off between benefits of homes in community with different values of $P_{mid}(t)$.}
\label{trade_off}
\end{figure}

Fig. \ref{trade_off} depicts the daily energy cost of each home in the community with the proposed CEMS under three values of variable $P_{mid}(t)$. It is clear that a trade-off occurs between the benefits of the homes that have and do not have a PV system in a community that applies MMR pricing. The benefits of the homes that have a PV system are increased and the benefits of the homes that do not have a PV system are decreased steadily when variable $P_{mid}(t)$ is increased. 

\section{Discussion}
One problem of a system-centric CEMS is the poor computational time for solving its optimization problem. However, in this study, using advanced mathematical solvers in AIMMS software, we demonstrate that the optimization problem of a community with the proposed CEMS can be transformed and solved within a short time even for a large community with 500 homes. Moreover, compared to the prosumer-centric CEMS, the benefit that the system-centric CEMS provides to the community is significant: the daily energy cost of the community with the proposed CEMS is only half that of the community with the prosumer-centric CEMS. In the system-centric CEMS, a central operation unit can flexibly increase or decrease the benefits of different types of homes depending on the policy of the community by using local energy trading with MMR pricing. This is a significant advantage of CB-LEM compared to P2P-LEM. Owing to these advantages, the system-centric CEMS and CB-LEM still exhibit enormous potential for future studies.

Other important problems of the system-centric CEMS are how to protect the private information of prosumers and a single point of failure at the central operation unit of the community. However, these problems are beyond the scope of this study.

\section{Conclusions and Future Works}
In this study, an optimization problem for a system-centric CEMS that supports different types of homes has been constructed and solved. Our proposed CEMS also supports local energy trading between homes using MMR pricing. A case study of a community in Detroit city, USA was used as a real scenario for testing the performance and computational time of the proposed CEMS. Moreover, we established two different CEMSs for the community for comparison: a prosumer-centric CEMS and no CEMS. The numerical results demonstrate that the daily energy cost of the proposed CEMS is the best among the three CEMSs and the computational time of the proposed CEMS for a 500-home community is only $118.2 s$, which is a sufficiently short time for day-ahead scheduling. Furthermore, with local energy trading using MMR pricing, the benefits of several homes can be changed flexibly by adjusting the value of variable $P_{mid}(t)$.

The security for a system-centric CEMS may be a potential field for future studies. Another research direction is to develop a cooperative strategy for many communities, whereby a community can buy or sell energy from another community.


%





\ifCLASSOPTIONcaptionsoff
  \newpage
\fi



\bibliographystyle{IEEEtran}
\bibliography{IEEEabrv,my_references}
\end{document}